\documentstyle[twoside]{article}

\catcode`\@=11
\long\def\@makefntext#1{
\protect\noindent \hbox to 3.2pt {\hskip-.9pt  
$^{{\eightrm\@thefnmark}}$\hfil}#1\hfill}		

\def\@makefnmark{\hbox to 0pt{$^{\@thefnmark}$\hss}}	
	
\def\ps@myheadings{\let\@mkboth\@gobbletwo
\def\@oddhead{\hbox{}
\rightmark\hfil\eightrm\thepage}   
\def\@oddfoot{}\def\@evenhead{\eightrm\thepage\hfil
\leftmark\hbox{}}\def\@evenfoot{}
\def\sectionmark##1{}\def\subsectionmark##1{}}

\oddsidemargin=\evensidemargin
\newcounter{sectionc}\newcounter{subsectionc}\newcounter{subsubsectionc}
\renewcommand{\section}[1] {\vspace{12pt}\addtocounter{sectionc}{1} 
\setcounter{subsectionc}{0}\setcounter{subsubsectionc}{0}\noindent 
	{\tenbf\thesectionc. #1}\par\vspace{5pt}}
\renewcommand{\subsection}[1] {\vspace{12pt}\addtocounter{subsectionc}{1} 
	\setcounter{subsubsectionc}{0}\noindent 
	{\bf\thesectionc.\thesubsectionc. {\kern1pt \bfit #1}}\par\vspace{5pt}}
\renewcommand{\subsubsection}[1] {\vspace{12pt}\addtocounter{subsubsectionc}{1}
	\noindent{\tenrm\thesectionc.\thesubsectionc.\thesubsubsectionc.
	{\kern1pt \tenit #1}}\par\vspace{5pt}}
\newcommand{\nonumsection}[1] {\vspace{12pt}\noindent{\tenbf #1}
	\par\vspace{5pt}}

\newcommand{\textlineskip}{\baselineskip=13pt}
\newcommand{\smalllineskip}{\baselineskip=10pt}

\def\eightcirc{
\begin{picture}(0,0)
\put(4.4,1.8){\circle{6.5}}
\end{picture}}
\def\eightcopyright{\eightcirc\kern2.7pt\hbox{\eightrm c}} 

\newcommand{\copyrightheading}[1]
	{\vspace*{-2.5cm}\smalllineskip{\flushleft
        {\footnotesize Nuovo Cimento B 114 (May 1999) 569-574 #1}\\
        {\footnotesize Los Alamos electronic archives: quant-ph/9707044 #1}\\
        {\footnotesize $\eightcopyright$\,
        1999 Nuovo Cimento \& H.C. Rosu
        }\\
	 }}

\def\abstracts#1#2#3{{
	\centering{\begin{minipage}{4.5in}\baselineskip=10pt\footnotesize
	\parindent=0pt #1\par 
	\parindent=15pt #2\par
	\parindent=15pt #3
	\end{minipage}}\par}} 

\renewenvironment{thebibliography}[1]
	{\frenchspacing
	 \ninerm\baselineskip=11pt
	 \begin{list}{\arabic{enumi}.}
        {\usecounter{enumi}\setlength{\parsep}{0pt}     
	 \setlength{\leftmargin 12.7pt}{\rightmargin 0pt} 
         \setlength{\itemsep}{0pt} \settowidth
	{\labelwidth}{#1.}\sloppy}}{\end{list}}

\newcounter{itemlistc}
\newcounter{romanlistc}
\newcounter{alphlistc}
\newcounter{arabiclistc}

\def\@citex[#1]#2{\if@filesw\immediate\write\@auxout
	{\string\citation{#2}}\fi
\def\@citea{}\@cite{\@for\@citeb:=#2\do
	{\@citea\def\@citea{,}\@ifundefined
	{b@\@citeb}{{\bf ?}\@warning
	{Citation `\@citeb' on page \thepage \space undefined}}
	{\csname b@\@citeb\endcsname}}}{#1}}

\newif\if@cghi
\def\cite{\@cghitrue\@ifnextchar [{\@tempswatrue
	\@citex}{\@tempswafalse\@citex[]}}
\def\citelow{\@cghifalse\@ifnextchar [{\@tempswatrue
	\@citex}{\@tempswafalse\@citex[]}}
\def\@cite#1#2{{$\null^{#1}$\if@tempswa\typeout
	{IJCGA warning: optional citation argument 
	ignored: `#2'} \fi}}

\def\@refcitex[#1]#2{\if@filesw\immediate\write\@auxout
	{\string\citation{#2}}\fi
\def\@citea{}\@refcite{\@for\@citeb:=#2\do
	{\@citea\def\@citea{, }\@ifundefined
	{b@\@citeb}{{\bf ?}\@warning
	{Citation `\@citeb' on page \thepage \space undefined}}
	\hbox{\csname b@\@citeb\endcsname}}}{#1}}

\def\@refcite#1#2{{#1\if@tempswa\typeout
        {IJCGA warning: optional citation argument
	ignored: `#2'} \fi}}

\def\refcite{\@ifnextchar[{\@tempswatrue
	\@refcitex}{\@tempswafalse\@refcitex[]}}

\def\pmb#1{\setbox0=\hbox{#1}
	\kern-.025em\copy0\kern-\wd0
	\kern.05em\copy0\kern-\wd0
	\kern-.025em\raise.0433em\box0}

\def\fnt#1#2{\footnotetext{\kern-.3em
	{$^{\mbox{\scriptsize #1}}$}{#2}}}

\def\runninghead#1#2{\pagestyle{myheadings}
\markboth{{\protect\footnotesize\it{\quad #1}}\hfill}
{\hfill{\protect\footnotesize\it{#2\quad}}}}
\font\tenrm=cmr10
\font\tenit=cmti10 
\font\tenbf=cmbx10
\font\bfit=cmbxti10 at 10pt
\font\ninerm=cmr9

\font\eightrm=cmr8

\def\qed{\hbox{${\vcenter{\vbox{			
   \hrule height 0.4pt\hbox{\vrule width 0.4pt height 6pt
   \kern5pt\vrule width 0.4pt}\hrule height 0.4pt}}}$}}

\begin{document}

\runninghead{Rosu and Romero, Ermakov approach
$\ldots$} {Rosu and Romero, 1D Helmholtz Hamiltonian
$\ldots$}


\normalsize\textlineskip
\thispagestyle{empty}
\setcounter{page}{1}

\copyrightheading{}			

\vspace*{0.88truein}

\centerline{\bf ERMAKOV APPROACH FOR THE ONE-DIMENSIONAL HELMHOLTZ
HAMILTONIAN
     }
\vspace*{0.035truein}
\vspace*{0.37truein}
\centerline{\footnotesize H. ROSU}
\vspace*{0.015truein}
\centerline{\footnotesize\it Instituto de F\'{\i}sica,
Universidad de Guanajuato, Apdo Postal E-143, 37150 Le\'on,
Guanajuato, Mexico}
\vspace*{10pt}
\centerline{\footnotesize J.L. ROMERO}
\vspace*{0.015truein}
\centerline{\footnotesize\it Departamento de F\'{\i}sica,
Universidad de Guadalajara, Corregidora 500, 44100 Guadalajara,
Jalisco, Mexico}
\vspace*{0.225truein}

\vspace*{0.21truein}
\abstracts{{\bf Summary}: -
For the one-dimensional Helmholtz equation we write the corresponding
time-dependent Helmholtz Hamiltonian in order to study it as an Ermakov
problem and derive geometrical angles and phases in this context.
}{}{}


\textlineskip                  
\vspace*{12pt}                 

\vspace*{1pt}\textlineskip	
\vspace*{-0.5pt}
\noindent


\noindent





\noindent
Our starting point is the Helmholtz equation in the form written by
Goyal {\em et al} [\refcite{1}] and by Delgado {\em et al} [\refcite{2}]
\begin{equation} \label{1}
\frac{d^2\psi}{dx^2}+\lambda \phi(x)\psi (x) =0~,
\end{equation}
i.e., as a Sturm-Liouville problem for the set of eigenvalues $\lambda \in R$
defining the Helmholtz spectrum on a given, closed interval [a,b] of the real
axis on which a non trivial $\psi$ is vanishing at both ends
(Dirichlet boundary conditions). Eq.~(1) occurs for example
in the case of TE (transverse electric) mode propagation in planar waveguides
with a refractive index which varies continuously in
the $x$ direction but is independent of $y$ and $z$ and in equivalent problems
in acoustics.
We perform the mapping of
Eq.~(1) to the canonical equations for a classical point
particle in the known way (for recent discussions see [\refcite{2,3}]).
Let $\psi (x)$ be any real solution of Eq.~(1). Defining x=t,
$\psi =q$, and $\psi ^{'}=p$, Eq.~(1) becomes
\begin{eqnarray}
\frac{dq}{dt}&=&p~\\
\frac{dp}{dt}&=&-\lambda\phi(t)q~,
\end{eqnarray}
with the spectral condition $q(a)=q(b)=0$. Eqs.~(2,3) are the canonical
equations of motion for a classical point particle as derived from the
time-dependent Hamiltonian
\begin{equation} \label{4}
H(t)=\frac{p^2}{2}+\lambda\phi(t)\frac{q^2}{2}~.
\end{equation}
For this Hamiltonian the triplet of phase-space
functions $T_1=\frac{p^2}{2}$, $T_2=pq$,
and $T_3=\frac{q^2}{2}$ forms a dynamical Lie algebra (i.e.,
$H=\sum _{n}h_{n}(t)T_{n}(p,q)$) which is closed with
respect to the Poisson bracket, or more exactly
$\{T_1,T_2\}=-2T_1$, $\{T_2,T_3\}=-2T_3$, $\{T_1,T_3\}=-T_2$. The Helmholtz
Hamiltonian can be written down as $H=T_1+\lambda \phi(t)T_3$. The so-called
Ermakov invariant $I$ [\refcite{erm}] belongs to the dynamical algebra
\begin{equation} \label{5}
I=\sum _{r}\mu _{r}(t)T_{r}~,
\end{equation}
and by means of
\begin{equation} \label{6}
\frac{\partial I}{\partial t}=-\{I,H\}~,
\end{equation}
one is led to the following equations for the unknown $\mu _{r}(t)$
\begin{equation} \label{7}
\dot{\mu} _{r}+\sum _{n}\Bigg[\sum _{m}C_{nm}^{r}h_{m}(t)\Bigg]\mu _{n}=0~,
\end{equation}
where $C_{nm}^{r}$ are the structure constants of the Lie algebra that
have been already given above. Thus, we get
\begin{eqnarray} \nonumber
\dot{\mu} _1&=&-2\mu _2 \\
\dot{\mu} _2&=&\lambda \phi(t)\mu _1-\mu _3\\
\dot{\mu} _3&=&2\lambda \phi(t)\mu _2~.    \nonumber
\end{eqnarray}
The solution of this system can be readily obtained by setting
$\mu _1=\rho ^2$ giving $\mu _2=-\rho \dot{\rho}$ and $\mu _3=\dot{\rho} ^2+
\frac{1}{\rho ^2}$, where $\rho$ is the solution of the Milne-Pinney equation
\cite{mp}
\begin{equation} \label{9}
\ddot{\rho}+\lambda \phi(t)\rho=\frac{1}{\rho ^3}~,
\end{equation}
In terms of the function $\rho (t)$ the Ermakov invariant can be written as
follows [\refcite{l}]
\begin{equation} \label{10}
I=\frac{(\rho p-\dot{\rho}q)^2}{2}+\frac{q^2}{2\rho ^2}~.
\end{equation}
Next, we calculate the time-dependent generating function allowing one to
pass to new canonical variables for which $I$ is chosen to be the
new ``momentum"
\begin{equation} \label{11}
S(q,I,\vec{\mu}(t))=\int ^{q}dq^{'}p(q^{'},I,\vec{\mu}(t))~,
\end{equation}
leading to
\begin{equation} \label{12}
S(q,I,\vec{\mu}(t))=\frac{q^2}{2}\frac{\dot{\rho}}{\rho}+
I{\rm arctan}\Bigg[\frac{q}{\sqrt{2I\rho ^2-q^2}}\Bigg]+
\frac{q\sqrt{2I\rho ^2-q^2}}{2\rho ^2}~, 
\end{equation}
where we have put to zero the constant of integration.
Thus
\begin{equation} \label{13}
\theta=\frac{\partial S}{\partial I}={\rm arctan}
\Big(\frac{q}{\sqrt{2I\rho ^2-q^2}}\Big)~.
\end{equation}
Moreover, the canonical variables are now
\begin{equation} \label{14}
q=\rho \sqrt{2I}\sin \theta ~,
\end{equation}
and
\begin{equation}  \label{15}
p=\frac{\sqrt{2I}}{\rho}\Big(\cos \theta+\dot{\rho}\rho\sin \theta\Big)~.
\end{equation}
The dynamical angle will be
\begin{equation} \label{16}
\Delta \theta ^{d}=\int _{t_0}^{t}\Bigg[\frac{1}{\rho ^2}-\frac{\rho ^2}{2}
\frac{d}{dt^{'}}\Big(\frac{\dot{\rho}}{\rho}\Big)\Bigg]dt^{'}
\end{equation}
whereas the geometrical angle reads
\begin{equation}  \label{17}
\Delta \theta ^{g}=\frac{1}{2}\int _{t_0}^{t}
\Bigg[(\ddot{\rho}\rho)-\dot{\rho}^2\Bigg] dt^{'}~.
\end{equation}
For periodic parameters $\vec{\mu}(t)$, with all the components of the same
period $T$, the
geometric angle is known as the (nonadiabatic) Hannay angle [\refcite{book}]
and in terms of $\rho$ is given by
\begin{equation}  \label{17b}
\Delta \theta ^{g}_{H}=-\oint _{C}\dot{\rho}d\rho~.
\end{equation}

\noindent
Passing now to the quantum Ermakov problem we turn $q$
and $p$ into quantum-mechanical operators, $\hat{q}$ and
$\hat{p}=-i\hbar\frac{\partial}{\partial q}$, but keeping the auxiliary
function $\rho$ as a $c$ number. The Ermakov invariant is a
constant Hermitian operator
\begin{equation}  \label{18}
\frac{dI}{dt}=\frac{\partial I}{\partial t}+\frac{1}{i\hbar}[I,H]=0
\end{equation}
and the time-dependent Schr\"odinger equation for the Helmholtz Hamiltonian
reads
\begin{equation}  \label{19}
i\hbar\frac{\partial}{\partial t}|\psi (\hat{q},t)\rangle=
\frac{1}{2}(\hat{p}^2+\lambda\phi(t)\hat{q}^2)|\psi(\hat{q},t)\rangle ~.
\end{equation}

\noindent
The goal now is to find the constant eigenvalues of $I$
\begin{equation}  \label{20}
I|\psi _{n}(\hat{q},t\rangle=\kappa _{n}|\psi _{n}(\hat{q},t)\rangle
\end{equation}
and to write the explicit superposition form of the general solution of
Eq.~(20) [\refcite{lr}]
\begin{equation}  \label{21}
\psi(\hat{q},t)=\sum _{n}C_{n}e^{i\alpha _{n}(t)}\psi _{n}(\hat{q},t)
\end{equation}
where $C_{n}$ are superposition constants, $\psi _{n}$ are the (orthonormal)
eigenfunctions of $I$, and the phases $\alpha _{n}(t)$, known as the
Lewis phases in the literature [\refcite{mor,m}], are to be found
from the equation
\begin{equation}  \label{22}
\hbar \frac{d\alpha _{n}(t)}{dt}=\langle \psi _{n}|i\hbar
\frac{\partial}{\partial t}-H|\psi _{n}\rangle~.
\end{equation}
The key point for the quantum Ermakov problem is to perform a unitary
transformation in order to obtain a transformed
eigenvalue problem for the new Ermakov invariant $I^{'}=UIU^{\dagger}$
possessing time-independent eigenvalues.
It is easy to get the required unitary transformation
as $U=\exp [-\frac{i}{\hbar}\frac{\dot{\rho}}{\rho}\frac{\hat{q}^2}{2}]$ and
the new Ermakov invariant will be $I^{'}=\frac{\rho ^{2}\hat{p}^2}{2}+
\frac{\hat{q}^{2}}{2\rho ^{2}}$. Therefore, its eigenfunctions are
$\propto e^{-\frac{\theta ^2}{2\hbar}}H_{n}(\theta/\sqrt{\hbar})$, where
$H_{n}$ are Hermite polynomials, $\theta=\frac{q}{\rho}$, and the eigenvalues
are $\kappa _{n}=\hbar(n+\frac{1}{2})$.
Thus, one can write the eigenfunctions $\psi _{n}$ as follows
\begin{equation}  \label{23}
\psi _{n}\propto \frac{1}{\rho ^{\frac{1}{2}}}
\exp \Big(\frac{1}{2}\frac{i}{\hbar}
\frac{\dot{\rho}}{\rho}q^2\Big)\exp\Big(-\frac{q^2}{2\hbar \rho ^2}\Big)
H_{n}\Big(\frac{1}{\sqrt{\hbar}}\frac{q}{\rho}\Big)~.
\end{equation}
The factor $1/\rho ^{1/2}$ has been introduced for normalization reasons.
Using these functions and performing
simple calculations one is lead to the geometrical phase
\begin{equation}  \label{23b}
\alpha _{n}^{g}=-\frac{1}{2}(n+\frac{1}{2})\int _{t_0}^{t}
\Bigg[(\ddot{\rho}\rho)-\dot{\rho}^2\Bigg] dt^{'}~.
\end{equation}
The cyclic (nonadiabatic) Berry phase [\refcite{book}] reads
\begin{equation}  \label{23c}
\alpha _{B,n}^{g}=(n+\frac{1}{2})\oint _{C}\dot{\rho}d\rho~.
\end{equation}

\noindent
Our last issue is the adiabatic approximation. We have found that a good
adiabaticity parameter is the inverse square root of the Helmholtz
eigenvalue,
$\frac{1}{\sqrt{\lambda}}$, leading to a ``slow time"
variable $\tau=\frac{1}{\sqrt{\lambda}} t$.
The adiabatic approximation has been thoroughly studied by
Lewis [\refcite{l}].
If the Helmholtz Hamiltonian is written as
\begin{equation} \label{24}
H(t)=\frac{\sqrt{\lambda}}{2}[p^2+\phi(t)q^2]~,
\end{equation}
then the corresponding Milne-Pinney equation reads
\begin{equation} \label{25}
\frac{1}{\lambda}\ddot{\rho}+\phi(t)\rho=\frac{1}{\rho ^3}~,
\end{equation}
while the Ermakov invariant will be a $1/\sqrt{\lambda}$-dependent function
\begin{equation} \label{26}
I(1/\sqrt{\lambda})=
\frac{(\rho p-\dot{\rho}q/\sqrt{\lambda})^2}{2}+\frac{q^2}{2\rho ^2}~.
\end{equation}
In the adiabatic approximation, Lewis [\refcite{l}] obtained the general
solution
of the Milne-Pinney equation in terms of the linearly independent solutions
$f$ and $g$ of the equation of motion $\frac{1}{\lambda}
\ddot{q}+\Omega ^2(t)q$
for the classical oscillator (see Eq. (45) in [\refcite{l}]).
Among the explicit examples given by Lewis, it is
$\Omega(t)=bt^{m/2}$, $m\neq-2$, $b={\rm const}$,
which is directly related to the dielectric
planar waveguides since it corresponds to power-law index profiles
($n(x)\propto x^{m/2}$).
For this case, Lewis obtained a simple formula for the $\rho$ function when
required to be $O(1)$ in $1/\sqrt{\lambda}$
\begin{equation}   \label{27}
\rho _{m}=\gamma _1\Bigg[\frac{\gamma _2\pi\sqrt{\lambda}}
{(m+2)}\Bigg]^{\frac{1}{2}}
t^{\frac{1}{2}}[H_{\beta}^{(1)}(y)H_{\beta}^{(2)}(y)]^{\frac{1}{2}}~,
\end{equation}
where $H_{\beta}^{(1)}$ and $H_{\beta}^{(2)}$ are Hankel functions of order
$\beta =1/(m+2)$,
$y=\frac{2b\sqrt{\lambda}}{(m+2)}t^{\frac{m}{2}+1}$, and $\gamma _1=\pm 1$,
$\gamma _2=\pm 1$.
Even more useful for the technological application may be
Lewis's specialization
to $m=-\frac{4n}{2n+1}$, $n=\pm 1,\pm 2, ...$ leading to
\begin{equation}  \label{28}
\rho _{n}=
\gamma _1\gamma _2^{\frac{1}{2}}b ^{-\frac{1}{2}}t^{\frac{n}{2n+1}}
|G(t,1/\sqrt{\lambda})|^2~,
\end{equation}
where
\begin{equation}  \label{29}
G(t,1/\sqrt{\lambda})=
\Bigg[\sum _{k=0}^{n}(-1)^{k}\frac{(n+k)!}{k!(n-k)!}
\Big(\frac{1/\sqrt{\lambda}}
{2ib(2n+1)}\Big)^{k}t^{-\frac{k}{(2n+1)}}\Bigg]^{\frac{1}{2}}~.
\end{equation}
Thus, one gets $\rho$ as a polynomial in the square of the
adiabaticity parameter, i.e.
$\lambda ^{-1}$, which has an infinite radius of convergence.
One can calculate the topological quantities by plugging the above
explicit Milne-Pinney functions into the corresponding formulas.
However, Lewis [\refcite{l}] carried out and checked a recursive formula
in $1/\lambda$ to order $1/\lambda ^3$, which can be used for any
index profile. It is
\begin{equation}  \label{30}
\rho = \rho _{0}+\rho _{1}/\lambda +\rho _{2}/\lambda ^2 +
\rho _{3}/\lambda ^3+...~,
\end{equation}
with $\rho _{0}=\Omega ^{-1/2}=\phi ^{-1/4}(x)$; for the other
coefficients $\rho _{i}$
see the appendix in [\refcite{l}]. The main contribution to the topological
quantities
comes from $\rho _{0}$. In the case of power-law index profiles one obtains
the geometrical angle
\begin{equation}  \label{31}
\Delta \theta ^{g}=-\frac{m}{4b(m+2)}\Bigg[t^{-(\frac{m}{2}+1)} -
t_{0}^{-(\frac{m}{2}+1)}\Bigg]
\end{equation}
and a similar formula for the geometrical quantum phase. For periodic
index profiles, one can work out Hannay angle and Berry's phase according
to their cycle integrals.
Finally, we notice that choosing $\phi (x)=\Phi (x)
+ \frac{{\rm Const}}{\psi ^3(x)}$,
which means nonlinear waveguides, leads to a somewhat more general
time-dependent Hamiltonian that has been discussed in the Ermakov perspective
by Maamache [\refcite{m}].

\noindent
In summary, we presented an application of the Ermakov procedure to
one-dimensional Helmholtz problems in a way which may be of direct experimental
relevance. We notice that very recently Petrov [\refcite{pet}] discussed the
evolution of Berry's phase in a graded-index medium, whereas several years
ago, Goncharenko {\em et al} [\refcite{gon}] studied the Ermakov approach in
nonlinear optics of elliptic Gaussian beams.

\nonumsection{Acknowledgements}
\noindent
This work was partially supported by the CONACyT Project 458100-5-25844E.
The second author thanks the University of
Guanajuato for a ``Scientific Summer" grant.


\nonumsection{References}


\end{document}